\documentstyle[12pt]{article}
\begin{document}
\begin{titlepage}
\title{Self-Dual Yang-Mills Fields\\ in\\ Eight Dimensions}
\author{ Ay\c se H\"umeyra Bilge${}^*$, Tekin Dereli ${}^\dagger$,
\c Sahin Ko\c cak${}^*$ \\     \\
${}^*$   {\small Department of Mathematics,}\\ {\small Anadolu University,}\\
{\small  Eski\c{s}ehir, Turkey}\\
{\footnotesize e.mail: bilge@yunus.mam.tubitak.gov.tr}\\   \\
${} ^\dagger$  {\small Department of Mathematics,}\\
{\small Middle East Technical University,}\\
{\small Ankara, Turkey} \\
{\footnotesize e.mail: tdereli@rorqual.cc.metu.edu.tr}}
\date{ }
\maketitle
\begin{abstract}
Strongly self-dual Yang-Mills fields in even dimensional spaces
are characterised by a set of constraints   on
the eigenvalues of the Yang-Mills fields $F_{\mu \nu}$.
We derive  a topological bound  on ${\bf R}^8$,
$\int_{M} ( F,F )^2  \geq 
k \int_{M} p_1^2$  where $p_1$ is the first Pontrjagin class
of the SO(n) Yang-Mills bundle and $k$ is a constant.
Strongly self-dual Yang-Mills fields realise the lower bound.
\end{abstract}
\end{titlepage}
\newpage
\noindent
{\bf 1. Introduction.}
\vskip .3cm
Self-dual Yang-Mills fields in four dimensions have many
remarkable properties. But
the notion of self-duality cannot be easily generalised to higher dimensions.
Let us first highlight the main points concerning the self dual
Yang-Mills theory in four dimensions before we discuss
a generalisation to higher dimensions. Our notation and conventions are
given in the Appendix.

{1.}
In four dimensions the Yang-Mills functional is the $L_2$ norm of the
curvature 2-form $F$
\begin{equation}
\| F\|^2=\int_M {\rm tr} (F \wedge *F)
\end{equation}
where $*$ denotes the Hodge dual of $F$.
The Yang-Mills equations are the critical points of the
Yang-Mills functional. They are of the form
\begin{equation}
d_EF=0,\quad ^*d_E^*F=0,
\end{equation}
where $d_E$ is the bundle covariant derivative and
$-*d_E*$ is its formal adjoint.

 {2.}
 $F$ is self-dual or anti-self-dual provided
\begin{equation}
*F=\pm F.
\end{equation}
Self-dual or anti-self-dual solutions of (2) are the global extrema of the Yang-Mills
functional (1).  This can be seen as
follows:
The Yang-Mills functional has a topological bound
\begin{equation}
\|F\|^2 \geq \int_M {\rm tr} (F\wedge F).
\end{equation}
The term ${\rm tr} (F\wedge F)$ is related to the Chern classes
of the bundle ${}^{[1]}$. Actually if $E$ is a
complex 2-plane bundle with $ c_1(E)=0$,then
the topological bound is proportional to $ c_2(E)$ and
this lower bound is realised by a (anti)self-dual  connection.
Furthermore, $SU(2)$ bundles over a four manifold are classified by $\int
c_2(E)$, hence self-dual connections are minimal representatives of the
connections in each  equivalence class of $SU(2)$ bundles ${}^{[2]}$.
This is a generalisation of the fact that an $SU(2)$ bundle admits a flat
connection if and only if it is trivial.

{3.}
Self-dual Yang-Mills equations form an elliptic system in the Coulomb gauge
${\rm div} A=0$. The importance of
this property is that, in gauge theory, elliptic equations lead to finite
dimensional moduli spaces ${}^{[3]}$.

 {4.}
Hodge duality can equivalently be defined by requiring that $F$ is an
eigen bi-vector of the completely antisymmetric  $\epsilon_{ijkl}$
tensor. In four dimensions such a tensor is unique and have the same constant
coefficients in all coordinate systems. Thus self-dual Yang-Mills
equations are linear equations that have the same constant coefficients in all
coordinate systems that leave the Euclidean metric invariant .

 {5.}
Self-dual Yang-Mills equations can be written
in terms of the commutator of two linear operators
depending on a complex parameter. Then their solution can be reduced to the
solution of a Riemann problem and this case is considered as completely
solvable in  integrable system theory .

The earliest attempt at a generalisation of self-duality
to higher dimensions is due to Tchrakian${}^{[4]}$.
His approach is pursued further for physical applications${}^{[5]}$.
In a similar vein,
Corrigan,Devchand,Fairlie and Nuyts${}^{[6]}$
studied  the first-order  equations satisfied by  Yang-Mills  fields
in spaces of dimension
greater than four (property 4 above) and derived
$SO(7)$ self-duality equations in ${\bf R}^8$.
This construction was followed by an investigation of completely solvable
Yang-Mills equations (property 5 above) in higher dimensions by Ward ${}^{[7]}$.
A different approach is taken by Feza G\"{u}rsey and his coworkers ${}^{[8]}$
who showed that the notion
of self-duality can be carried over to eight dimensions
by making use of octonions.
This observation
proved  fruitful as $SO(7)$ self-duality  equations  in ${\bf R}^8$
can be  solved in terms of octonions ${}^{[9]}$ .
The existence of the so called {\em octonionic instantons}
is closely tied to the properties of the Cayley algebra. These
turned out to be relevant to superstring models.  In fact
soliton solutions of the string equations of motion
to the lowest order in the string tension parameter,
that preserve at least one supersymmetry, are given by
octonionic instantons ${}^{[10,11]}$.
A further connection between strings and octonions is provided
by the existence of supersymmetric Yang-Mills theories and
superstring actions only in dimensions 3,4,6,10. The number of
transverse dimensions 1,2,4,8  coincide with the dimensions
of the division algebras {\bf R},{\bf C}, {\bf H} and {\bf O}${}^{[12]}$.
These are related with the four Hopf fibrations, the last of which
$ S^{15} \rightarrow S^8$ is realised by the octonionic instantons ${}^{[13]}$.

Here we  present a characterisation of strongly self-dual
Yang-Mills fields by an eigenvalue criterion
in even dimensional spaces.
In particular, the self-dual Yang-Mills
equations in four dimensions and
the self-duality equations of Corrigan et al in eight dimensions are consistent
with our strong self-duality criteria.

\vskip 3mm
\noindent {\bf 2. Self-duality of a 2-form as an eigenvalue criterion.}
\vskip 4mm
Any 2-form in $2m$ dimensions can be represented
by a $2m\times 2m$
skew symmetric real matrix. Such a matrix $\Omega$
has pure imaginary eigenvalues that are
pairwise conjugate. Let  $\lambda_k$, $k=1,\dots m$ be these
eigenvalues. This is equivalent to the existence of a basis $\{X_k$, $Y_k\}$
such that
\begin{equation}
\Omega X_k =-\lambda_k Y_k,\quad \Omega Y =\lambda_k X_k,
\end{equation}
for each $k$.  In this basis the matrix $\Omega$ takes the block diagonal form:
\begin{equation}
\Omega = \left( \begin{array}{ccccccc}
                          0&\lambda_1&&&&& \\
                         -\lambda_1&0&&&&& \\
                           &      &.&&&& \\
                           &      &&.&&& \\
                           &      &&&.&& \\
                           &      &&&&0&\lambda_m \\
                           &      &&&&-\lambda_m&0
                            \end{array} \right)
\end{equation}

We first consider the case $m=2$.
In four dimensions self-duality can be described by the following
eigenvalue criterion
which allows an immediate generalisation to higher dimensions.
Let $\Omega$ be the skew symmetric
matrix representing a 2-form $F$ in four dimensions. Then it can be seen that
the eigenvalues of the matrix $\Omega$ satisfy
\begin{equation}
\lambda_1\mp\lambda_2=\sqrt{(\Omega_{12}\mp \Omega_{34})^2+
(\Omega_{13}\pm \Omega_{24})^2
         +(\Omega_{14}\mp \Omega_{23})^2}
\end{equation}
Thus for self-duality
\begin{equation}
\lambda_1= \lambda_2,
\end{equation}
while for anti-self-duality
\begin{equation}
\lambda_1= -\lambda_2.
\end{equation}
In both cases the absolute values of the
eigenvalues are equal. Two cases are distinguished by the sign of
\begin{equation}
\Omega_{12}\Omega_{34}-\Omega_{13}\Omega_{24}+
\Omega_{14}\Omega_{23}.
\end{equation}

In higher dimensions we propose the following

\vskip 3mm
\noindent
{\bf Definition }
Let $F$ be a  2-form
in $2m$ dimensional space and $\pm i\lambda_k$, $k=1,\dots,m$ be its
eigenvalues.  Then $F$ is {\it strongly self-dual} if it can be brought to block
diagonal form (6) with respect to an oriented basis such that
$\lambda_1=\lambda_2=\dots=\lambda_m$.
\vskip 3mm
\noindent The strong self-duality
or strong anti-self-duality
can be characterised by requiring the equality of the absolute values
of the eigenvalues:
\begin{equation}
|\lambda_{1}| = |\lambda_{2}| = \dots  = |\lambda_{m}| .
\end{equation}
There are $2^m$ possible ways to satisfy the above set
of equalities. The half of these correspond to strongly self-dual
2-forms,  while  the  remaining  half  correspond  to  strongly  anti-self-dual
2-forms.

Now  consider  a 2-form $F$ in eight dimensions and let the corresponding
matrix be $\Omega$. Then
\begin{equation}
{\rm det}(I+t\Omega)= 1+\sigma_2 t^2+\sigma_4 t^4
+\sigma_6 t^6+\sigma_8 t^8
\end{equation}
where, provided the eigenvalues of $\Omega$ are
$\quad \lambda_1, \lambda_2, \lambda_3, \lambda_4,\quad$
we have
\begin{eqnarray}
\sigma_2&=&\lambda_1^2+\lambda_2^2+\lambda_3^2+\lambda_4^2 = tr\Omega^2
\nonumber \\
\sigma_4&=&\lambda_1^2\lambda_2^2+\lambda_1^2\lambda_3^2
                +\lambda_1^2\lambda_4^2 +\lambda_2^2\lambda_3^2
                +\lambda_2^2\lambda_4^2 +\lambda_3^2\lambda_4^2
\nonumber \\
\sigma_6&=&\lambda_1^2\lambda_2^2\lambda_3^2
                +\lambda_1^2\lambda_2^2 \lambda_4^2
                +\lambda_1^2\lambda_3^2 \lambda_4^2
                +\lambda_2^2\lambda_3^2 \lambda_4^2 \nonumber \\
\sigma_8&=&\lambda_1^2\lambda_2^2 \lambda_3^2\lambda_4^2 = Det\Omega.
\end{eqnarray}

The sign of square root of the determinant
\begin{equation}
\sqrt{\sigma_8} =\lambda_1 \lambda_2  \lambda_3 \lambda_4
\end{equation}
distinguishes between the self-dual and anti-self-dual cases.

In the next section we will need various estimates involving the norm of
$F$. To this end we  prove the following

\proclaim Lemma 1. Let $F \in \Lambda^2({\bf R}^8)$. Then
$3  (F, F)^2-2 (F^2, F^2) \ge 0$

\noindent
{\it Proof.} From above it can be seen that
\begin{eqnarray}
3 (F,F)^2-2 (F^2, F^2) &=&3 \sigma_2^2
-8 \sigma_4 \nonumber \\
&=&3(\lambda_1^4+
\lambda_2^4+\lambda_3^4+\lambda_4^4) \nonumber \\
& &-2(
\lambda_1^2\lambda_2^2+\lambda_1^2\lambda_3^2
                +\lambda_1^2\lambda_4^2 +\lambda_2^2\lambda_3^2
                +\lambda_2^2\lambda_4^2 +\lambda_3^2\lambda_4^2)
\nonumber \\
&=&(\lambda_1^2-\lambda_2^2)^2
 +(\lambda_1^2-\lambda_3^2)^2
 +(\lambda_1^2-\lambda_4^2)^2 \nonumber \\
& &+(\lambda_2^2-\lambda_3^2)^2
 +(\lambda_2^2-\lambda_4^2)^2
 +(\lambda_3^2-\lambda_4^2)^2 \nonumber \\
&\ge&0.\quad\quad   \Box   \nonumber
\end{eqnarray}

\proclaim Lemma 2. Let $F \in \Lambda^2({\bf R}^8)$. Then
$( F^2,F^2) - *F^4 \ge 0$

{\it Proof.} From above we have
\begin{eqnarray}
{1\over4}(( F^2,F^2) - *F^4)&=&(
\sigma_4-6\sqrt{\sigma_8})\nonumber \\
&=&(\lambda_1\lambda_2-\lambda_3\lambda_4)^2
 +(\lambda_1\lambda_3-\lambda_2\lambda_4)^2
 +(\lambda_1\lambda_4-\lambda_2\lambda_3)^2 \nonumber \\
 &\ge&0\quad. \quad \Box \nonumber
\end{eqnarray}

Equality holds in both the above lemmas if and only if
the absolute values of all eigenvalues are equal.

The computation above allows us also to prove the following

\proclaim Proposition. Let $ F \in \Lambda^2({\bf R}^8)$ and
$F_{\mu \nu}$ be its coordinate components. If $\pm \lambda_k$,
$k=1,\dots, 4$ are the corresponding eigenvalues, then
$\lambda_1=\lambda_2=\lambda_3=\lambda_4$ if the
following equations  are satisfied:
\begin{eqnarray}
&&F_{12}-F_{34}=0\quad F_{12}-F_{56}=0\quad F_{12}-F_{78}=0 \nonumber \\
&&F_{13}+F_{24}=0\quad F_{13}-F_{57}=0\quad F_{13}+F_{68}=0 \nonumber \\
&&F_{14}-F_{23}=0\quad F_{14}+F_{67}=0\quad F_{14}+F_{58}=0 \nonumber \\
&&F_{15}+F_{26}=0\quad F_{15}+F_{37}=0\quad F_{15}-F_{48}=0 \nonumber \\
&&F_{16}-F_{25}=0\quad F_{16}-F_{38}=0\quad F_{16}-F_{47}=0 \nonumber \\
&&F_{17}+F_{28}=0\quad F_{17}-F_{35}=0\quad F_{17}+F_{46}=0 \nonumber \\
&&F_{18}-F_{27}=0\quad F_{18}+F_{36}=0\quad F_{18}+F_{45}=0
\end{eqnarray}

{\it Proof.} From the proof of Lemma 1, $3\sigma_2^2-8\sigma_4=0$
if and only if
$\lambda_1=\lambda_2=\lambda_3=\lambda_4$. On the other hand, by direct
computation with the EXCALC package in REDUCE it is shown that this condition
holds when the above equations  are satisfied.  $\Box$

These are precisely the self-duality equations of
Corrigan,Devchand, Fairlie and Nuyts ${}^{[6]}$.
\vskip 4mm
\noindent {\bf 3. A topological lower bound }
\vskip 3mm

How to choose the action density and its topological lower bound?
The idea underlying the self-dual Yang-Mills theory is that the Yang-Mills
action $\|F\|^2$ (that is a second order curvature invariant) is
bounded below by a topological invariant of the bundle.
In the case of an $SU(2)$ bundle over a four manifold,
the bundle is classified
by the second Chern number $\int c_2(E)$. This lower bound is achieved
for self-dual
fields, hence a self-dual connection on an $SU(2)$ bundle is the connection
whose norm is minimal among all connections on that bundle.

In higher dimensions, we want similarly to find a lower bound for the norm of
a curvature invariant in terms of topological invariants of the bundle.
If the base manifold is eight dimensional, we will need to integrate an 8-form,
hence assuming that the first Chern class of the bundle vanishes $c_1(E)=0$ ,
an appropriate lower bound can be given by a linear combination of the
only available Pontryagin classes :
\begin{equation}
\beta p_1(E)^2+\gamma p_2(E).
\end{equation}
But then the topological lower bound would be a fourth order expression in $F$,
so that it is natural to seek a lower bound to some fourth order curvature
invariant.
The action density should be written in terms of the
local curvature 2-form
matrix in a way independent of the local trivilization of the bundle. Hence it
should involve invariant polynomials of the local curvature matrix. We want to
express the action as an inner product in the space of $k$-forms, which gives a
quantity independent of the local coordinates. This leads to
the following generic action density
\begin{equation}
a ( F,F )^2+b  ( F^2,F^2 )
+ c (trF^2,trF^2) .
\end{equation}

The action density proposed by Grossman,Kephardt and Stasheff ${}^{[13]}$
corresponds to the choice $a=c=0$. Then
the self-duality (in the Hodge sense) of $F^2$ gives global minima of this
action involving the characteristic number    $\int p_2(E)$.

We derive an alternative lower bound involving the other
characteristic number $\int p_{1}(E)^{2}$.
In analogy with the fact that in four dimensions the $L_2$ norm of the
curvature is bounded below by a topological number,
we give a lower bound for the
expression $\int ( F,F ) ^2$ for an $SO(n)$
bundle. However, from the Schwarz inequality,
it  can be seen that for a compact manifold M,
\begin{equation}
\int_M (F,F) \le \big(\int (F, F)^2\big)^{1/2}{\rm vol}M,
\end{equation}
hence lower bounds on the $\int ( F,F )^2$
cannot be related to the $L_2$ norm of $F$.

Consider an $SO(n)$ valued curvature 2-form matrix $F=(F_{ab})$. Then
\begin{equation}
( F, F ) = 2\sum_{a=1}^n\sum_{b>a}^n(F_{ab},F_{ab})
\end {equation}
and
\begin{eqnarray}
( F,F)^2&=&4\sum_{a=1}^n\sum_{b>a}(F_{ab},F_{ab})^2 \nonumber \\
& & +8\sum_{a=1}^n\sum_{b>a}\sum_{c>a}^n\sum_{d>c}(F_{ab},F_{ab})
(F_{cd},F_{cd})
\end{eqnarray}
On the other hand
\begin{equation}
{\rm tr}F^2=-2\sum_{a=1}\sum_{b>a}F_{ab}^2
\end{equation}
hence
\begin{eqnarray}
( {\rm tr}F^2,{\rm tr}F^2) &=&
4\sum_{a=1}^n\sum_{b>a}(F_{ab}^2,F_{ab}^2) \nonumber \\
& &+8\sum_{a=1}^n\sum_{b>a}\sum_{c>a}^n\sum_{d>c}(F_{ab}^2,F_{cd}^2).
\end{eqnarray}
From the Schwarz inequality we have
\begin{eqnarray}
({\rm tr}F^2,{\rm tr}F^2) &=&
4\sum_{a=1}^n\sum_{b>a}(F_{ab}^2,F_{ab}^2) \nonumber \\
& & +8\sum_{a=1}^n\sum_{b>a}\sum_{c>a}\sum_{d>c}(F_{ab}^2,F_{ab}^2)^{1/2}
(F_{cd}^2,F_{cd}^2)^{1/2} .
\end{eqnarray}
Then from Lemma 1 we obtain
\begin{equation}
( F,F) ^2 \ge   {2 \over 3} ({\rm tr}F^2,{\rm tr}F^2),
\end{equation}
while from Lemma 2
\begin{equation}
( trF^2,trF^2 ) \ge *tr F^4.
\end{equation}
In both (24) and (25) the lower bound is realised by
strongly (anti)self-dual fields.
\vskip 4mm
\noindent {\bf 4. Conclusion}
\vskip 4mm
In this paper we have characterised strongly (anti)self-dual
Yang-Mills fields in even dimensional spaces by putting constraints on the
eigenvalues of $F$. The previously known cases of self-dual
Yang-Mills fields in four and eight dimensions are consistent with our
characterisation.
We believe this new approach to self-duality in higher dimensions
deserves further study.

We have also derived a topological bound
\begin{equation}
\int (F,F)^2\ge k \int_{M}  p_1(E)^2
\end{equation}
that is achieved by strongly (anti)self-dual Yang-Mills fields.
This is to be compared with a topological  bound obtained
earlier ${}^{[13]}$
\begin{equation}
\int_{M} (F^2 ,F^2)  \ge k' \int_M p_{2}(E).
\end{equation}
As  already noted,
the self-duality of $F^2$ in the Hodge sense attains the lower bound in (27).
But as opposed to the self duality of $F$ in four dimensions, this
condition is too restrictive since it yields an overdetermined system for
the connection , namely 21 nonlinear  equations for  8 variables.

The new topological bound we found here will also be important for applications
because, for instance, there are bundles  such that
$\int p_2(E) =0$ while $ \int p_1(E)^2 \neq 0$.${}^{[14]}$
Given any sequence of numbers $p_{1},p_{2},...,p_{n},$ they possibly
cannot be realised as Pontryagin numbers of a 4n-manifold.
But it can be shown that for an appropriate $k$, the multiples $kp_{1},
kp_{2},...,kp_{n}$ can be realised. As a corollary, there are infinitely 
many distinct 8=manifolds with $p_{2} = 0$ but $p_{1} \neq 0$.

\newpage

\noindent {\bf Appendix:  Notation and Conventions}
\vskip 2mm
Let $M$ be an $m$ dimensional differentiable
manifold, and $E$ be a vector bundle over
$M$ with standard fiber $R^n$ and  structure group $G$.
A connection on $E$
can be represented by $\cal G$ valued connection 1-form $A$,
where $\cal G$ is a
linear representation of the Lie algebra of  the structure group $G$. Then the
curvature $F$ of the connection $A$ can be represented
locally by a $\cal G$ valued 2-form given by
\begin{equation}
F=dA-A\wedge A .
\end{equation}
We omit the wedge symbols in what follows.  The local curvature
2-form depends on a trivilization of the bundle, but its invariant
polynomials $\sigma_k$ defined by
\begin{equation}
{\rm det}(I+tF)=\sum_{k=0}^n\sigma_k t^k
\end{equation}
are invariants of the local trivialization. We recall that  $\sigma_k$ 's
are closed $2k$-forms,
hence they define  the
deRham cohomology classes in $H^{2k}$. Furthermore these
cohomology
classes depend only on the bundle.  They are  the Chern classes of the
bundle $E$ up to some multiplicative constants [10].
 It is also known that the $2k$-form $\sigma_k$ can be obtained as
linear combinations of $trF^k$ where $F^k$ means the product of the matrix $F$
with itself $k$ times, with the wedge multiplication of the entries.

$(A,B)$ denotes the inner product in the fibre
where $A, B$ are Lie algebra valued
forms. In particular if  $\alpha,\beta$ are scalar valued p-forms on $M$
then $(\alpha,\beta) = \alpha \wedge * \beta$.
The $L_2$ norm squared of $A$ on the manifold
$||A||^2 = \int_M (A,A) d{\rm vol}$.

The indices $(\mu,\nu\dots)$ will be used as space-time
indices while the indices $(a,b,\dots)$  for Lie algebra basis.
Let $e^\mu$, $\mu=1,\dots,8$  be any orthonormal basis of 1-forms
$\Lambda^1(R^8)$. Then
$e^{\mu\nu}=e^\mu e^\nu$, $\mu<\nu$ will denote the basis of 2-forms.
Similarly, $e^{\mu\nu\alpha\beta}$, $\mu>\nu>\alpha>\beta$ will be a basis of
4-forms. Let $E_a$ be a basis for the Lie algebra of the structure
group. Then we can write $F$ as
\begin{equation}
F= \sum_{\mu>\nu}F_{\mu\nu}e^{\mu\nu}=
 \sum_{a>b} F^{ab} E_{ab} =
\sum_{a>b}\sum_{\mu>\nu}F^{ab}_{\mu\nu}
E_{ab} \otimes e^{\mu\nu}.
\end{equation}
Note that $F_{\mu\nu}$ is an element of the Lie algebra, $F^{ab}$ are 2-forms
and $F^{ab}_{\mu\nu}$ are functions on $M$.
\newpage
{\bf References}
\vskip 4mm
\begin{description}
\item [[1]] J.M.Milnor and J.D. Stasheff, {\bf Characteristic Classes}
(Princeton U.P., 1974)
\item [[2]] C.H.Taubes, Contemp. Math. {\bf 35}(1984)493 and references therein.
\item [[3]] S.K.Donaldson and P.B.Kronheimer {\bf The Geometry of Four
Manifolds} (Clarendon , Oxford, 1990)
\item [[4]] D. H. Tchrakian, J. Math. Phys. {\bf 21}(1980)166
\item [[5]] T.N.Sherry, D.H.Tchrakian, Phys.Lett.{\bf B147}(1984)121
\\ D.H.Tchrakian, Phys.Lett.{\bf B150}(1985)360
\\ C. Sa\c{c}l{\i}o\u{g}lu, Nucl.Phys.{\bf B277}(1986)487
\item [[6]] E.Corrigan, C.Devchand, D.B.Fairlie, J.Nuyts,
Nucl.Phys.{\bf B214} 452(1983)452
\item [[7]] R.S.Ward , Nucl.Phys.{\bf B236}(1984)381
\item [[8]] R.D\"{u}ndarer,F.G\"{u}rsey,C.-H.Tze, J. Math.Phys.{\bf 25}(1984)1496
\item [[9]] D.B.Fairlie,J.Nuyts, J. Phys.{\bf A17}(1984)2867
\\  S.Fubini,H.Nicolai,Phys. Lett.{\bf 155B}(1985)369
\\ R.D\"{u}ndarer,F.G\"{u}rsey,C.-H.Tze,
 Nucl. Phys.{\bf B266}(1986)440
\item [[10]] J.A.Harvey,A.Strominger, Phys. Rev. Lett.{\bf 66}(1991)549
\item [[11]] T.A.Ivanova, Phys. Lett. {\bf B315}(1993)277
\item [[12]] J. M. Evans, Nucl. Phys. {\bf B298}(1988)92
\item [[13]] B.Grossman, T.W.Kephart, J.D.Stasheff, Commun. Math.
Phys.{\bf 96}(1984)431  Err. ,ibid {\bf 100}(1985) 311
\item [[14]] F. Hirzebruch {\bf Topological Methods in Algebraic Geometry}\\
(Springer, 1966)

\end{description}
\end{document}